\documentclass[12pt]{article}

\title{\bf Two and three electrons\\
           in a quantum dot: $1/|J|$ - expansion}
\author{Augusto Gonzalez$^{1,2}$, Ricardo Perez$^2$ and 
        Pavel Fileviez$^2$}
\date{$^1$Depto. de Fisica, Univ. Nacional de Colombia,\\ 
      Sede Medellin, AA 3840, Medellin, Colombia\\
      and\\
      $^2$Instituto de Cibernetica, Matematica y Fisica,\\
      Calle E 309, Vedado, Habana 4, Cuba}
\begin{document}

\maketitle

\begin{abstract}
We consider systems of two and three electrons in a two-dimensional
parabolic quantum dot. A magnetic field is applied perpendicularly 
to the electron plane of motion. We show that the energy levels
corresponding to states with high angular momentum, $J$, and a low 
number of vibrational quanta may be systematically computed as power
series in $1/|J|$. These states are relevant in the high-$B$ limit.
\end{abstract}
\newpage

\section{Introduction}

Recently, few-electron systems in nearly two-dimensional semiconductor 
structures has been a subject of intense theoretical and experimental 
research [1]. In the present paper, we continue a programme, initiated 
in [2, 3], aimed at providing reliable analytical estimates for the 
energy levels of model two-dimensional systems on the basis of a 
simple physical picture for the states.

The idea is to use the inverse of angular momentum as an expansion 
parameter in Schrodinger equation, thus obtaining nonperturbative 
series for the energy and the wave function. Variants of this method 
have been widely used in Atomic and Nuclear Physics [4, 5].                                                

The static exciton in a magnetic field was considered in [2]. Energy 
levels were computed as a function of the mass ratio and the magnetic 
field. Comparison with estimates obtained from two-point Pade 
approximants [6] yielded excellent results at any magnetic field
strength.

In paper [3], a model of three electrons in a quantum dot with $1/r^2$ 
repulsion, considered previously in [7], was studied. The 
$1/|J|$-method provided analytical estimates for the energy and a set 
of approximate quantum numbers to label the states. We will make use 
below of many of the results obtained in [3].

Systems of two and three electrons in a two-dimensional parabolic 
quantum dot are to be studied in the present paper. A magnetic field 
is applied perpendicularly to the dot's plane. In this problem, there 
is only one relevant parameter, $\beta=(E_{coul}/\hbar\Omega)^{1/6}$, 
where $E_{coul}$ is the characteristic Coulomb energy, and $\Omega$ 
is expressed in terms of the dot and the cyclotron frequencies as 
$\Omega=\sqrt{\omega_0^2+\omega_c^2/4}$. $\beta\to 0$ will be called 
the oscillator limit, and $\beta\to\infty$ -- the Wigner limit, 
suggesting that we are dealing with a few-electron version of the 
Wigner solid. In typical quantum dot experiments, $\beta\sim 1$.

The physical picture emerging from the $1/|J|$-expansion is the 
following. In the leading approximation, the electrons are located 
at the vertices of a regular polygon, which side minimises the total 
classical energy (centrifugal plus potential). The first corrections 
take account of small (harmonic) vibrations around the equilibrium 
configuration. Higher corrections come from anharmonic oscillations. 
This picture is supported by numerical calculations for few-electron 
systems not only in the strong coupling regime [8], but in the region 
$\beta\sim 1$ as well [9].          

A special comment deserves paper [9], in which the few-electron 
problem in Eckardt coordinates is treated quasiclassically. Although 
the starting points are far apart, our results are very similar up to 
the harmonic approximation. The main difference between paper [9] and 
ours is the expansion parameter, which is $\hbar$ in [9], and $1/|J|$ 
in our work. Higher order of the quasiclassical expansion are not 
reported in [9].

The plan of the paper is as follows. In section 2, we present the 
$1/|J|$-method. Results for two and three electrons are shown in 
sections 3 and 4. Concluding remarks are given in the last section.

\section{The $1/|J|$-expansion}

We start with the dimensionless hamiltonian governing the 
two-dimensional motion of $N$ electrons in a quantum dot of energy 
$\hbar\omega_0$, and in a magnetic field normal to the plane of 
motion,

\begin{equation}
\frac{H}{\hbar\Omega} = \frac{1}{2}\sum_{k=1}^N\left(p_k^2+r_k^2
    \right) + \beta^3 \sum_{k<l} \frac{1}{r_{kl}}
    + \frac{\omega_c}{2 \Omega} \sum_{k=1}^N J_k
    + \frac{g \omega_c}{2 \Omega} S_z.
\end{equation}

The conventions are as follows. $\mu$ is the electron effective mass, 
$\omega_c=eB/\mu c$ is the cyclotron frequency, $g$ is the effective 
gyromagnetic factor, $S_z$ is the $z$ component of the total spin of 
the system, $\Omega=\sqrt{\omega_0^2+\omega_c^2/4}$ is the effective 
dot frequency. The length unit is $\sqrt{\hbar/(\mu\Omega)}$. The 
parameter $\beta$ is given by 
$\beta^3=\sqrt{\frac{\mu e^4}{\kappa^2\hbar^2}/\hbar\Omega}$, 
where $\kappa$ denotes the dielectric constant.

The actual parameter governing the problem is $\beta$. When 
$\beta<<1$, the levels are small perturbations around oscillator 
energies. On the other hand, when $\beta \to \infty$, the energy 
minimum is reached in a classical configuration which can be seen as 
a few-body model of a Wigner crystal. Both limiting situations 
conduce to exactly solvable problems. 

Through the introduction of Jacobi coordinates, the center-of-mass 
motion is separated from the internal motion,

\begin{equation}
\frac{H}{\hbar\Omega} = \frac{H_{cm}}{\hbar\Omega} 
     + \frac{H_{int}}{\hbar\Omega},
\end{equation}

\noindent
where

\begin{eqnarray}
\frac{H_{cm}}{\hbar\Omega} &=& -   
   \left(\frac{\partial^2}{\partial\rho_{cm}^2}
   +\frac{1}{\rho_{cm}}\frac{\partial}{\partial\rho_{cm}} \right)
   +\frac{J_{cm}^2}{\rho_{cm}^2} + \frac{1}{4} \rho_{cm}^2
   +\frac{\omega_c}{2\Omega} J_{cm},\\
\frac{H_{int}}{\hbar\Omega} &=& -\sum_{k=1}^{N-1} \left(
          \frac{\partial^2}{\partial\rho_k^2} + \frac{1}{\rho_k}
          \frac{\partial}{\partial\rho_k} \right)
          - \sum_{k=1}^{N-2} \left(                     
          \frac{1}{\rho_k^2}+\frac{1}{\rho_{k+1}^2}
          \right) \frac{\partial^2}{\partial\theta_k^2}\nonumber\\
&+& 2 \sum_{k=1}^{N-3} \frac{1}{\rho_{k+1}^2}
          \frac{\partial^2}{\partial\theta_k\partial\theta_{k+1}} 
      + \frac{2 i J}{N-1} \sum_{k=1}^{N-2} \left(\frac{1}{\rho_k^2}
 - \frac {1}{\rho_{k+1}^2} \right)\frac{\partial}{\partial\theta_k}
          \nonumber\\
&+& \sum_{k=1}^{N-1} \left( \frac{J^2}{(N-1)^2}
          \frac{1}{\rho_k^2} + \frac{1}{4} \rho_k^2 \right)
          + \beta^3 \sum_{k<l} \frac{1}{r_{kl}} 
          + \frac{\omega_c}{2 \Omega} J 
          + g \frac{\omega_c}{2 \Omega} S_z,
\end{eqnarray}

\noindent
We have written $H_{int}$ in terms of rotational invariant 
coordinates, i.e., the moduli of the Jacobi vectors and the angles 
between them. The Jacobi vectors were defined as,

\begin{equation}
\vec \rho_k = \sqrt{\mu_k/\mu_1} \left\{ \vec r_{k+1} - \frac{1}{k}
          \sum_{l=1}^k \vec r_l \right\},~~~~~~~k = 1, 2, ..., N-1,
\end{equation}

\noindent
where the reduced masses are $\mu_k=k/(k+1)$. The angle between 
$\vec\rho_k$ and $\vec\rho_{k+1}$ is denoted $\theta_k$. $J$ labels 
the total internal angular momentum.

The eigenvalues of $H_{cm}$ are simply,

\begin{equation}
\frac{E_{cm}}{\hbar\Omega} = 2 n_{cm} + |J_{cm}| + 1
     + \frac{\omega_c}{2 \Omega} J_{cm}.
\end{equation}

We shall obtain approximate expressions for the eigenvalues of 
$H_{int}$. We consider states with high $|J|$ values, which are the 
relevant states at high magnetic fields. It is intuitively evident 
that at high $|J|$, $E_{int}\sim |J|$, $<\rho_k^2> \sim |J|$. The 
solution of the Schrodinger equation may be organised as a power 
series in $1/|J|$. A scaling of dimensions such that 
$\rho^2 \to |J| \rho^2$ makes evident the dependence of each term on 
$|J|$. The scaled hamiltonian is written as

\begin{eqnarray}
h &=& \frac{1}{|J|} \left\{ \frac{H_{int}}{\hbar\Omega}
      - \frac{\omega_c}{2 \Omega} J 
      - \frac{g \omega_c}{2 \Omega} S_z \right\}\nonumber\\
  &=& \sum_{k=1}^{N-1} \left\{ \frac{1}{(N-1)^2}
      \frac{1}{\rho_k^2} + \frac{1}{4} \rho_k^2 \right\}
       + \tilde\beta^3 \sum_{k<l} \frac{1}{r_{kl}}\nonumber \\
  &+& \frac{1}{J^2} \left\{- \sum_{k=1}^{N-1} \left(            
      \frac{\partial^2}{\partial\rho_k^2} 
      + \frac{1}{\rho_k} \frac{\partial}{\partial\rho_k} \right)
      - \sum_{k=1}^{N-2} \left(                         
      \frac{1}{\rho_k^2}+\frac{1}{\rho_{k+1}^2}\right)        
      \frac{\partial^2}{\partial\theta_k^2}\right.\nonumber\\
  &+& \left. 2 \sum_{k=1}^{N-3} \frac{1}{\rho_{k+1}^2}
      \frac{\partial^2}{\partial\theta_k\partial\theta_{k+1}}
      \right\}+\frac{2 i}{J(N-1)} \sum_{k=1}^{N-2} \left(
      \frac{1}{\rho_k^2}-\frac {1}{\rho_{k+1}^2} \right)
      \frac{\partial} {\partial\theta_k}, 
\end{eqnarray}

\noindent
where we have introduced the ``renormalised'' coupling constant 
$\tilde\beta^3 = \beta^3/|J|^{3/2}$. When taking the formal limit 
$|J| \to \infty$, $\tilde\beta^3$ is kept fixed to take account of 
Coulomb repulsion nonperturbatively.

In what follows, we consider only the two- and three-electron 
problems. The only term surviving in the r.h.s of (7) when 
$|J| \to \infty$ is the effective potential. Its absolute minimum 
gives a classical contribution to the energy. It is reached in a 
configuration where the particles sit at the corners of a regular 
polygon. In this configuration, the effective potential is a function 
of $\rho_1$ only, 

\begin{equation}
U_{eff} = \frac{2 \sin^2 \pi/N}{N \rho_1^2} +
     \frac {N \rho_1^2}{8 \sin^2 \pi/N} + 
     \frac {\tilde\beta^3 \sin \pi/N}
     {\rho_1} \sum_{i<j} \frac{1}{|\sin \theta_{ij}/2|},
\end{equation}

\noindent
where $\theta_{ij}$ is the angle between particles $i$ and $j$, 
measured from the c.m. Minimisation of $U_{eff}$ leads to an 
equilibrium value for $\rho_1$, $\rho_{10}$. The equilibrium values 
of the others coordinates, $\rho_{k0}$, $\theta_{k0}$, are determined 
from geometry.

Higher contributions to the energy come from relaxing the equilibrium 
configuration, $\rho_k = \rho_{k0} + y_k/|J|^{1/2}$, $\theta_k = 
\theta_{k0} + z_k/|J|^{1/2}$. The r.h.s. of (7) may, thus, be expanded 
as

\begin{equation}
h = h_0 + \frac{h_2}{|J|} + \frac{h_3}{|J|^{3/2}} + \frac{h_4}{|J|^2} 
        + ...,
\end{equation}

\noindent
where $h_0 = U_{eff}(\rho_{10})$, $h_2$ describes in general harmonic 
oscillations in fictitious magnetic fields, and $h_3$, $h_4$, etc 
account for anharmonicities. 

For the spatial wave function and the scaled energy (related to $h$), 
we write series like (9)

\begin{eqnarray}
\psi &=& \psi_0 + \frac{\psi_1}{|J|^{1/2}} + \frac{\psi_2}{|J|} 
         + ..., \\ 
\epsilon &=& \epsilon_0 + \frac{\epsilon_2}{|J|} + 
           \frac{\epsilon_4}{J^2} + \dots~~.
\end{eqnarray}

These expressions are substituted into Schrodinger equation, leading 
to the chain of uncoupled equations, 

\begin{eqnarray}
h_0 &=& \epsilon_0 \\
h_2 \psi_0 &=& \epsilon_2 \psi_0,~~~{\rm etc.}
\end{eqnarray}

\noindent
Higher corrections to the energy are obtained from ordinary 
perturbation theory, where $h_3$, $h_4$, etc. are interpreted as 
perturbations.

Below, we present results for two and three electrons.

\section{Two electrons}

For two particles, there is only one Jacobi coordinate, $\vec\rho_1 = 
\vec r_2 - \vec r_1$. It's equilibrium value (the modulus) satisfies 
the equation

\begin{equation}
0 = - 2 + \frac{1}{2} \rho_{10}^4 - \tilde\beta^3 \rho_{10}.
\end{equation}

Writing $\rho_1 = \rho_{10} + y_1/|J|^{1/2}$, and substituting into 
the r.h.s. of (7), we obtain the operator coefficients $h_2$, $h_3$, 
etc. In particular

\begin{eqnarray}
h_2 &=& - \frac{{\rm d}^2}{{\rm d} y_1^2} + 
         \frac{1}{2}~ U''(\rho_{10}) y_1^2, \nonumber \\
    &=& \omega_1 (a_1^{\dagger} a_1 + \frac{1}{2}),  
\end{eqnarray}

\noindent
where $\omega_1=\sqrt{2 U''(\rho_{10})}=\sqrt{3+4/\rho_{10}^4}$, 
$a_1=\omega^{1/2} y_1/2 + ip_1/\omega^{1/2}$, $a_1^{\dagger}=
\omega^{1/2} y_1/2 - ip_1/\omega^{1/2}$. Note that in the oscillator 
limit $\omega_1 \to 2$, whereas in the Wigner limit $\omega_1$ 
approaches the classical result $\sqrt{3}$ [10].
 
To this order, the spatial wave function is written as $e^{i J \theta} 
\psi_0 (y_1)$, where $\psi_0 \sim (a_1^{\dagger})^n |0>$, and $\theta$ 
is the polar angle associated to the vector $\vec \rho_1$. Under a 
permutation of particles $\theta$ changes by $\pi$, thus even $|J|$ 
correspond to unpolarised spin states, $S=0$, and odd $|J|$ -- to 
polarised states, $S=1$.

The operators $h_k$, $k \ge 3$, can be written in general as

\begin{equation}
h_k = \frac{(-1)^k}{\rho_{10}^{k-2}} \left\{ \left( 
      \frac{k-1}{\rho_{10}^4} + \frac{1}{2} \right) y_1^k 
      + y_1^{k-3} \frac{\partial}{\partial y_1}\right\}.
\end{equation}

\noindent
Their matrix elements can be straightforwardly computed. Let us write 
the results for the first nonzero terms of the series (11) in the 
present case,

\begin{eqnarray}
\epsilon_0 &=& \frac{3}{4} \rho_{10}^2 - \frac{1}{\rho_{10}^2},\\
\epsilon_2 &=& \omega_1 (n + \frac{1}{2}),\\
\epsilon_4 &=& \frac{3}{\rho_{10}^2                
               \omega_1^2}\left(\frac{3}{\rho_{10}^4}
               + \frac{1}{2}\right)(2 n^2+2 n+1)-\frac{1}
               {4\rho_{10}^2}\nonumber\\
           &&  -\frac{1}{\rho_{10}^2                
               \omega_1^4}\left(\frac{2}{\rho_{10}^4}
               + \frac{1}{2}\right)^2(30 n^2+30 n+11),\\
\epsilon_6 &=& -\frac{15}{8 \rho_{10}^4 \omega_1^9} 
                (4/\rho_{10}^4 +1)^4
                (2 n+1)(47 n^2+47 n+31)\nonumber\\
           &&  +\frac{9}{2 \rho_{10}^4 \omega_1^7} 
                (4/\rho_{10}^4 +1)^2
                (6/\rho_{10}^4 +1) (2 n+1) (25 n^2+25 n+19)
                \nonumber\\
           &&  - \frac{1}{2 \rho_{10}^4 \omega_1^5} 
                 (6/\rho_{10}^4 +1)^2
                 (2 n+1)(87 n^2+87 n+86)\nonumber\\
           &&  + \frac{10}{\rho_{10}^{12} \omega_1^5} (2 n+1) 
                 (14 n^2+14 n+13)\nonumber\\
           &&  + \frac{1}{\rho_{10}^4 \omega_1^3}(2 n+1)
                 \{(10/\rho_{10}^4+1)
                 (5 n^2+5 n+9) - 9/\rho_{10}^4 \}\nonumber\\
           &&  - \frac{3}{4 \rho_{10}^4 \omega_1} (2 n+1),~~{\rm etc}.
\end{eqnarray}

$n$ labels the number of excitation quanta. It is a good approximate 
quantum number of the problem in the whole interval $0\le\beta<\infty$. 
Notice that all the $\epsilon_k$ with $k>2$ go to zero in both the 
oscillator and Wigner limits. Thus, the $1/|J|$-expansion provides a 
nice interpolation scheme.

In figure 1a, we compare different approximations to the r.h.s. of (7) 
computed according to the series (11). The excited state with $n=2$ 
and $|J|=3$ is shown. Convergence is excellent. We can not distinguish 
among the curves containing the corrections $\epsilon_4$ and 
$\epsilon_6$ (printed as dashed lines) from the solid line 
corresponding to $\epsilon_0+\epsilon_2/|J|$. Even at a low value of 
$|J|$ as 3, $\epsilon_4$ introduces corrections lower than 1\%, and 
$\epsilon_6$ -- lower than 0.3\%. The relative weight of $\epsilon_6$ 
in $\epsilon=\epsilon_0+\epsilon_2/|J|+\epsilon_4/J^2+\epsilon_6/|J|^3$
is shown in figure 1b.

In figure 2, the energy $|J| \epsilon$ is compared with 
the estimate obtained in [11] from a two-point Pade approximant. We 
expect this approximant to approach the exact energy from above, and 
to differ from it in no more than a few parts in $10^3$ [11]. The 
results are consistent with the expectations.

Notice that the conditions for the application of perturbation theory 
are fulfilled at any $\beta$ when $|J|\ge 3$, i.e. the distances 
between adjacent levels, computed from $\epsilon_0+\epsilon_2/|J|$, 
are much greater than the corrections $\epsilon_4/|J|^2+\dots$. Below, 
we will see that in the three-body problem the corrections, although 
being small, may cause a partial rearrangement of the spectrum. 

\section{Three electrons}

In the three electron system, there are two distances $\rho_1$, 
$\rho_2$, and one angle, $\theta_1$. The equilibrium configuration is 
an equilateral triangle ($\rho_{10}=\rho_{20}$, 
$\theta_{10}=\pm\pi/2$), which side satisfies the equation

\begin{equation}
0 = -1 + \rho_{10}^4 -3 \tilde\beta^3 \rho_{10}.
\end{equation}

At its minimum, $U_{eff}(\rho_{10})=3 \rho_{10}^2/2-\rho_{10}^{-2}/2$, 
thus providing a leading approximation to the energy

\begin{equation}
\epsilon_0 = \frac{3}{2} \rho_{10}^2 - \frac{1}{2 \rho_{10}^2}.
\end{equation}

Expanding around one of the two equivalent configurations, i.e. fixing 
$\theta_{10}=\pi/2$ for example, we obtain a series like (9) in which 
the first operator coefficients are given by

\begin{eqnarray}
h_2=&-&\left(\frac{\partial^2}{\partial y_1^2}+
             \frac{\partial^2}{\partial y_2^2}+
      \frac{2}{\rho_{10}^2}\frac{\partial^2}{\partial z_1^2}\right)
             -\frac{2 i}{\rho_{10}^3}{\rm sign}(J) (y_1-y_2)
             \frac{\partial}{\partial z_1}\nonumber\\
    &+&\frac{1}{4}\left(\frac{3}{\rho_{10}^4}+1\right)(y_1^2+y_2^2)
       \nonumber\\
    &+&\frac{1}{16}\left(1-\frac{1}{\rho_{10}^4}\right)
       (5 y_1^2+6 y_1 y_2+5 y_2^2+ 3\rho_{10}^2 z_1^2),
\end{eqnarray}

\begin{eqnarray}
h_3=&-&\frac{1}{\rho_{10}}\left(\frac{\partial}{\partial y_1}+
       \frac{\partial}{\partial y_2}\right)
       +\frac{2}{\rho_{10}^3}(y_1+y_2)
       \frac{\partial^2}{\partial z_1^2}\nonumber\\
    &+&\frac{3 i}{\rho_{10}^4}{\rm sign}(J)(y_1^2-y_2^2)
       \frac{\partial}{\partial z_1}-              
       \frac{1}{\rho_{10}^5}(y_1^3+y_2^3)\nonumber\\
    &-&\frac{1}{64\rho_{10}}\left(1-\frac{1}{\rho_{10}^4}\right)
       (19 y_1^3+3 y_1^2 y_2+33 y_1 y_2^2+9 y_2^3\nonumber\\
    &-&9\rho_{10}^2 y_1 z_1^2+21\rho_{10}^2 y_2 z_1^2),
\end{eqnarray}

\begin{eqnarray}
h_4 &=&\frac{1}{\rho_{10}^2}\left(y_1\frac{\partial}{\partial y_1}+
        y_2\frac{\partial}{\partial y_2}\right)-
        \frac{3}{\rho_{10}^4}(y_1^2+y_2^2)
        \frac{\partial^2}{\partial z_1^2}\nonumber\\
    &-&\frac{4 i}{\rho_{10}^5}{\rm sign}(J)(y_1^3-y_2^3)
  \frac{\partial}{\partial z_1}+\frac{5}{4\rho_{10}^6}
        (y_1^4+y_2^4)\nonumber\\
    &+&\frac{1}{256\rho_{10}^2}\left(1-\frac{1}{\rho_{10}^4}\right)
       \left(\frac{329}{4} y_1^4-25 y_1^3 y_2+
       \frac{123}{2} y_1^2 y_2^2+135 y_1 y_2^3+
       \frac{9}{4}y_2^4 \right.\nonumber\\
    &-&\left.\frac{99}{2}\rho_{10}^2 y_1^2 z_1^2+27\rho_{10}^2 y_1
       y_2 z_1^2+\frac{141}{2}\rho_{10}^2 y_2^2
       z_1^2+\frac{41}{4}\rho_{10}^4 z_1^4\right),~{\rm etc}
\end{eqnarray}

The hamiltonian $h_2$, which is to be taken as zeroth order 
hamiltonian, describes a harmonic oscillator of frequency 
$\omega_1=\sqrt{3+1/\rho_{10}^4}$ in the variable $y_s=(y_1+y_2)/
\sqrt{2}$, plus a combination of harmonic oscillators and a fictitious 
magnetic field in the variables $y_m=(y_1-y_2)/\sqrt{2}$ and 
$z_m=\rho_{10} z_1/\sqrt{2}$. Note that the ``magnetic field'' comes 
from the coupling between the angular momentum and the variable 
$\theta_1$. We may use a symmetric gauge to describe the fictitious 
field. The hamiltonian is transformed according to 
$h'=e^{i f} h e^{-i f}$, where $f={\rm sign}(J) y_m z_m/
(2 \rho_{10}^2)$. $h'_2$ takes the symmetric form

\begin{eqnarray}
h'_2=&-&\frac{\partial^2}{\partial y_s^2}+\frac{1}{4}
       \left(3+\frac{1}{\rho_{10}^4}\right)y_s^2
       -\left(\frac{\partial^2}{\partial \xi^2}+
       \frac{1}{\xi}\frac{\partial}{\partial \xi}
       +\frac{1}{\xi^2}\frac{\partial^2}{\partial\alpha^2}\right)
          \nonumber\\
    &-& i~\frac{{\rm sign}(J)}{\rho_{10}^2}\frac{\partial}{\partial
    \alpha}+\frac{1}{8}\left(3-\frac{1}{\rho_{10}^4}\right) \xi^2,
\end{eqnarray}

\noindent
where $\xi=\sqrt{y_m^2+z_m^2}$ and $\alpha=\arctan(z_m/y_m)$. The 
corresponding eigenvalues and eigenfunctions are

\begin{eqnarray}
\epsilon_2 &=& \omega_1 (n_s+1/2)+\omega_2 (2 n+|m|+1)
           +{\rm sign}(J) m \omega_3,\\
\psi_0 &\sim& H_{n_s}(\sqrt{\omega_1/2}y_s)
 e^{-\omega_1 y_s^2/4} \xi^{|m|}
 e^{-\omega_2\xi^2/4} L_n^{|m|}(\omega_2\xi^2/2) e^{i m \alpha},
\end{eqnarray}

\noindent
in which $\omega_2=\sqrt{3/2-\rho_{10}^{-4}/2}$, 
$\omega_3=1/\rho_{10}^2$, and $n$ and $m$ are respectively principal 
and magnetic quantum numbers in the fictitious field. In the 
oscillator limit: $\omega_1\to 2$, $\omega_2,~\omega_3\to 1$, whereas 
in the Wigner limit: $\rho_{10}\to\infty$, $\omega_1\to\sqrt{3}$, 
$\omega_2\to\sqrt{3/2}$, and $\omega_3\to 0$. At finite $\tilde\beta$ 
there is no degeneracy between the states of $h_2$, thus we may use 
nondegenerate perturbation theory to compute corrections.

We shall identify which of the levels ($J,n_s,n,m$) may describe 
electron states. This set of quantum numbers will serve as approximate 
quantum numbers for the three-electron problem. At intermediate 
couplings, a small mixing between the states of $h'_2$ will be induced 
by the corrections $h'_3$ and $h'_4$.

Let us recall the symmetry requirements for three-electron wave 
functions. A spatially antisymmetric wave function, 
$\Psi_0=e^{i J \Xi} \psi_0$, where $\Xi$ accounts for global rotations, 
corresponds to a spin-polarised state, $S=3/2$, whereas a 
mixed-symmetry $\Psi_0$ is related to an unpolarised spin state, 
$S=1/2$. Symmetry transformations shall be expressed in terms of the 
variables we are using, i.e. $y_s$, $\xi$ and $\alpha$. Additionally, 
to identify physical states, $\Psi_0$ shall be compared with oscillator 
functions at $\tilde\beta\to 0$. A detailed analysis is presented in 
the Appendix of paper [3], to which we refer the reader. The main 
conclusion is that in the $(n,m)$ plane, the spatially symmetric and 
antisymmetric states occupy the lines $m=J+3 k$ , where $k$ is an 
integer. For example, let us consider the first states with $J=-3$ 
when $\tilde\beta<<1$. The lowest state is ($n_s,n,m$)=(0,0,0). It 
corresponds to both one antisymmetric and one symmetric state. In our 
scheme, as tunnelling effects are not included, they are degenerated 
in energy. Of course, only the antisymmetric state may correspond to a 
state of three electrons. Very near to the lowest state there is the 
mixed-symmetry doublet, (0,0,1). At excitation energies near 2, there 
are the levels (1,0,0), (1,0,1), (0,0,-1), (0,1,0) and (0,1,1). 
Symmetric and antisymmetric states have $m=0$. The next set of states 
is at excitation energies near 4, etc.

Up to this order, the energy computed from $\epsilon=\epsilon_0+
\epsilon_2/|J|$ leads to results very similar to that reported in [9]. 
That is, when $\beta\sim 1$ the relative error of the lowest state with 
a given $|J|$ represents only a few percents of the total energy at 
$|J|=3$, and decreases considerably as $|J|$ is increased. The reader 
may look at figures 3 and 4 of paper [9] in quality of examples.

Let us study $\epsilon$ when $\beta$ is varied from 0 to $\infty$. As 
in the two-electron problem, the higher corrections, 
$\epsilon_4/|J|^2+\dots$, go to zero in both $\beta\to 0$ and 
$\beta\to\infty$ limits. However, for certain levels, a rearrangement 
of the spectrum may take place. For example, let us consider spatially 
antisymmetric states with $J=-3 k$, where $k$ is positive and high 
enough. The lowest state in this sector has quantum numbers (0,0,0). 
We show in figure 3 the excitation energies of three states with quantum 
numbers (0,0,3), (1,0,0) and (0,1,0). Their energies are 
$3(\omega_2-\omega_3)$, $\omega_1$ and $2 \omega_2$ respectively. 
Transitions  among these states induced by $h'_3$, $h'_4$, etc. may 
occur at $\tilde\beta\ge 0.6$, thus resulting in a rearrangement of 
levels. To compute their energies at the intermediate values of 
$\beta$ where the intersections occur, we shall resolve the degeneracy 
by means of degenerate perturbation theory. On the other hand, there 
are levels, in particular those with numbers (0,0,0), for which we can
apply the $1/|J|$-expansion continuously from $\tilde\beta=0$ to 
$\infty$.

Let us compute the coefficient $\epsilon_4$ for a state with quantum 
numbers (0,0,0). We assume $J<0$ and $|J|$ large enough. These are
the relevant states at high magnetic fields. We shall compute the 
matrix elements entering the expression

\begin{eqnarray}
\epsilon_4 &=& <0,0,0|h'_4|0,0,0>\nonumber\\
           &-& \sum_{n_s,n,m}
               \frac{<0,0,0|h'_3|n_s,n,m><n_s,n,m|h'_3|0,0,0>}
                    {n_s\omega_1+(2 n+|m|)\omega_2-m\omega_3}.
\end{eqnarray}

Taking into account the explicit form of $h'_3$, i.e.

\begin{equation}
h'_3=A \frac{\partial}{\partial y_s}+B y_s+C y_s^3+D, 
\end{equation}

\noindent
where

\begin{eqnarray}
A &=& -\frac{\sqrt{2}}{\rho_{10}}, \\
B &=& \frac{\sqrt{2}}{\rho_{10}}\left\{
      \frac{\partial^2}{\partial z_m^2}+ 
      2 i\frac{{\rm sign}(J)}{\rho_{10}^2} 
      y_m \frac{\partial}{\partial z_m}-\frac{1}{4\rho_{10}^4} y_m^2
      \right.\nonumber\\
  &&  \left. -\frac{3}{16}\left(1-\frac{1}{\rho_{10}^4}\right) 
      (y_m^2+z_m^2)\right\},\\
C &=& -\frac{\sqrt{2}}{4\rho_{10}}\left(1+\frac{1}{\rho_{10}^4}
      \right),\\
D &=& -\frac{5\sqrt{2}}{32\rho_{10}}\left(1-\frac{1}{\rho_{10}^4}
       \right)(y_m^3-3 y_m z_m^2),
\end{eqnarray}

\noindent
we obtain that the sum over intermediate states in (29) contains only
a few terms. The following result for $\epsilon_4$ arises,

\begin{eqnarray}
\epsilon_4 &=& <0,0,0|h'_4|0,0,0>-\frac{C^2}{3\omega_1}<3|y_s^3|0>^2
               \nonumber\\
           &-& \frac{1}{3}<0,3|D|0,0>^2 \left( 
               \frac{1}{\omega_2-\omega_3}+\frac{1}{\omega_2+\omega_3}
               \right)+\frac{A^2}{\omega_1}<1|
               \frac{\partial}{\partial y_s}|0>^2\nonumber\\      
           &-& \frac{2 C}{\omega_1}<1|y_s|0><1|y_s^3|0><0,0|B|0,0>
               \nonumber\\
           &-& \frac{C^2}{\omega_1}<1|y_s^3|0>^2
               -\frac{1}{\omega_1}<1|y_s|0>^2 <0,0|B|0,0>^2
               \nonumber\\
           &-& <1|y_s|0>^2 \left\{
               \frac{<0,2|B|0,0>^2}{\omega_1+2(\omega_2-\omega_3)}
               +\frac{<0,-2|B|0,0>^2}{\omega_1+2(\omega_2+\omega_3)} 
               \right\},
\end{eqnarray}

\noindent
where

\begin{eqnarray}
<0,0,0|h'_4|0,0,0> &=& -\frac{1}{\rho_{10}^2}
                       -\frac{3}{\omega_2 \rho_{10}^6}
                       \left(\frac{1}{\omega_2}+\frac{1}{\omega_1} 
                       \right)+ \frac{15}{8 \rho_{10}^6}\left(
                       \frac{1}{\omega_1}+\frac{1}{\omega_2} \right)^2 
                       \nonumber\\         
                   &+& \frac{3}{8 \rho_{10}^2}\left(1+
                       \frac{\omega_2}{\omega_1}+
                       \frac{1}{\omega_1 \omega_2 \rho_{10}^4}+ 
                       \frac{3}{\omega_2^2 \rho_{10}^4} \right)
                       \nonumber\\
                   &+& \frac{3}{64 \rho_{10}^2}
                       \left(1-\frac{1}{\rho_{10}^4}\right)
                       \left( \frac{13}{\omega_2^2}+ 
                       \frac{12}{\omega_1 \omega_2}+ 
                       \frac{16}{\omega_1^2} \right), \\
<0,3|D|0,0> &=& -\frac{5\sqrt{6}}{16\rho_{10} \omega_2^{3/2}}\left(
                1-\frac{1}{\rho_{10}^4}\right), \\
<0,0|B|0,0> &=& -\frac{\sqrt{2} \omega_2}{2 \rho_{10}}, \\
<0,\pm 2|B|0,0> &=& \pm \frac{1}{\rho_{10}^3}
                    -\frac{\omega_2}{4 \rho_{10}}
                    -\frac{1}{4\omega_2\rho_{10}^3},
\end{eqnarray}

\noindent
and the matrix elements involving $y_s$ are ordinary oscillator matrix 
elements, that is $<1|y_s|0>=1/\sqrt{\omega_1}$, etc. Note that, as 
mentioned, $\epsilon_4 \to 0$ in both the $\beta\to 0$ and 
$\beta\to\infty$ limits.

In figure 4 we compare the $1/|J|$-expansion and the two-point Pade 
approximant $P_{5,4}(\beta)$ of [11] for the lowest 
antisymmetric state with $J=-3$. As in the two-electron case, the 
relative error is of the same order of the incertitude of the Pade 
estimate, i.e. a few parts in $10^3$. We show in figure 5 the relative 
weight of $\epsilon_4$ in $\epsilon=\epsilon_0+\epsilon_2/|J|+
\epsilon_4/J^2$. For states with still higher values of $|J|$, the 
relative magnitude of the contribution of  $\epsilon_4$ decreases 
considerably.

\section{Conclusions}

We have computed the energy levels of two and three electrons in an 
ideal parabolic two-dimensional quantum dot in the presence of a 
magnetic field. States with high angular momentum and a low number of 
excitation (vibrational) quanta were considered. The energy was found 
as a series in the ``small'' parameter $1/|J|$. The series exhibits 
good convergence properties even at $|J|=3$.

At still lower values of the angular momentum the method is not
applicable. However, we can 
obtain reliable analytic estimations by means of an alternative 
procedure: the two-point Pade approximants, which are extremely simply 
constructed for the low-lying states [11]. $1/|J|$-expansions and Pade 
approximants were shown to give similar results at intermediate $|J|$ 
values, i.e. at $|J|=3$.

Other problems are currently being treated along these lines. For 
example, the effects of anyonic statistics on the energy levels of two 
quasiparticles in a quantum dot may be straightforwardly studied by 
means of the $1/|J|$-method [12].

\section{Acknowledgements}

One of the authors (A. G.) acknowledges support from the Colombian 
Institute for Science and Technology (COLCIENCIAS) under Project 1118 
- 05 - 661 - 95, and from the Associateship Scheme of the International 
Centre for Theoretical Physics (Trieste, Italy), where part of this 
work was done. 

\newpage

{\Large Figure Captions}
\vspace{1cm}

\begin{description}
\item{Figure 1.} 
  \begin{description}
         \item{a)} The energy of the $|J|=3$, $n=2$ state of two 
                   electrons.
         \item{b)} The relative weight of $\epsilon_6$ in $\epsilon$.
  \end{description}
\item{Figure 2.} Relative difference between $|J| \epsilon$ and the 
               $P_{6,5}$ Pade approximant for two electrons in a state 
               with $|J|=3$, $n=0$. 
\item{Figure 3.} Excitation energies of three states with $J=-3 k$,
               $k$ - positive and high enough. Quantum numbers are
               indicated.
\item{Figure 4.} Relative difference between $|J| \epsilon$ and the 
               $P_{5,4}$ Pade approximant for three electrons. The 
               lowest antisymmetric state with $J=-3$ is studied. 
\item{Figure 5.} Relative weight of $\epsilon_4$ in $\epsilon$. The same 
               state as in figure 4 is considered.
\end{description}


\newpage

\begin{thebibliography}{99}

\bibitem{} See, for example, Johnson N F 1995 J. Phys.: Cond. Matter 
           {\bf 7} 965, and references therein
\bibitem{} Quiroga L, Camacho A and Gonzalez A 1995 J. Phys.: Cond.
           Matter {\bf 7} 7517
\bibitem{} Gonzalez A, Quiroga L and Rodriguez B 1996 Few-Body
           Systems {\bf 21} 47
\bibitem{} Herschbach D R, Avery J and Goscinki O (eds.) 1993
           Dimensional Scaling in Chemical Physics: Kluwer Academic
\bibitem{} Gonzalez A 1991 Few-Body Systems {\bf 10} 43 
\bibitem{} Mac Donald A H and Ritchie D R 1986 Phys. Rev. B {\bf
           33} 8336
\bibitem{} Johnson N F and Quiroga L 1995 Phys. Rev. Lett. {\bf
           74}, 4277 
\bibitem{} Ruan W Y, Liu Y Y, Bao C G and Zhang Z Q 1995 Phys. Rev.
           B {\bf 51} 7942; Li X G, Ruan W Y, Bao C G and Liu Y Y,
           1997 Few-Body Systems, to appear
\bibitem{} Maksym P A 1996 Phys. Rev B {\bf 53} 10 871
\bibitem{} Schweigert V A and Peeters F M 1995 Phys. Rev. B {\bf
           51} 7700
\bibitem{} Gonzalez A 1997 J. Phys.: Cond. Matter, to appear
\bibitem{} Perez R and Gonzalez A, in preparation  

\end{thebibliography}
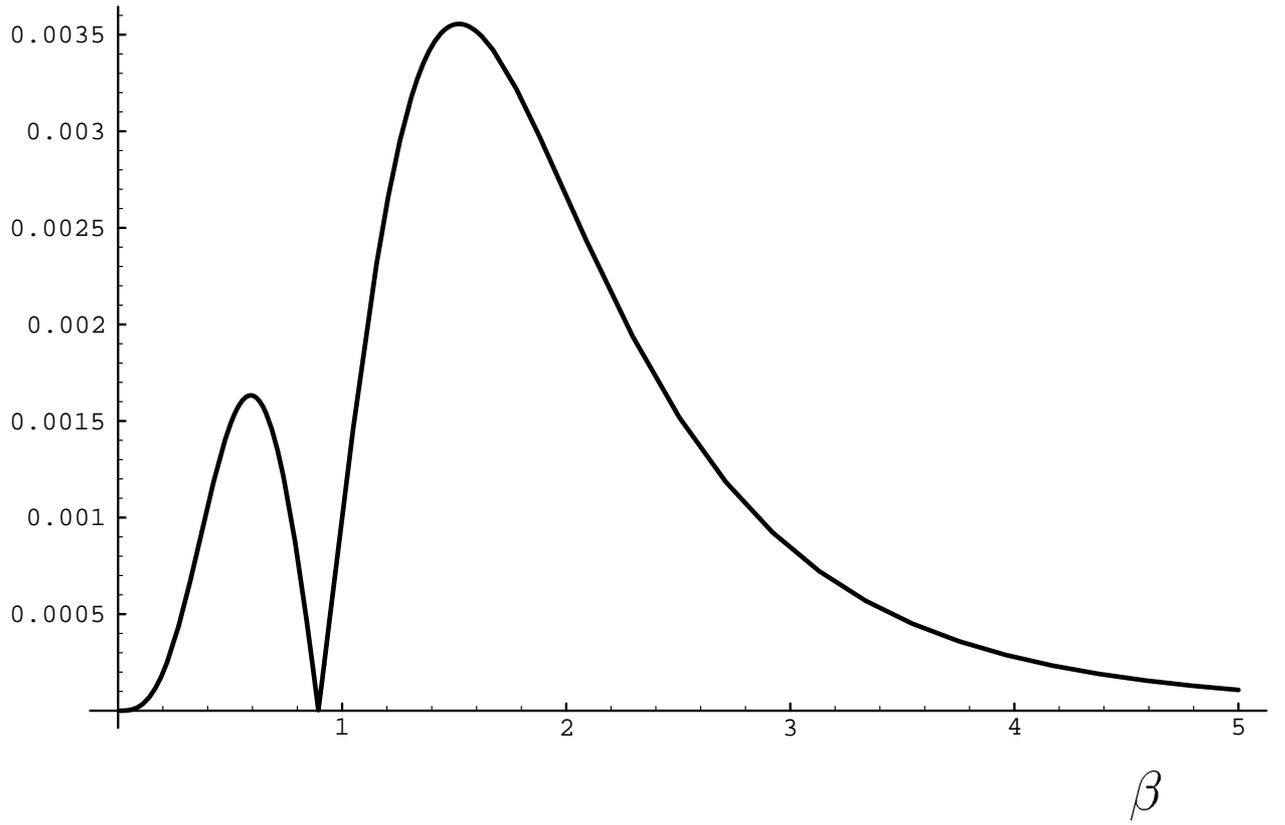
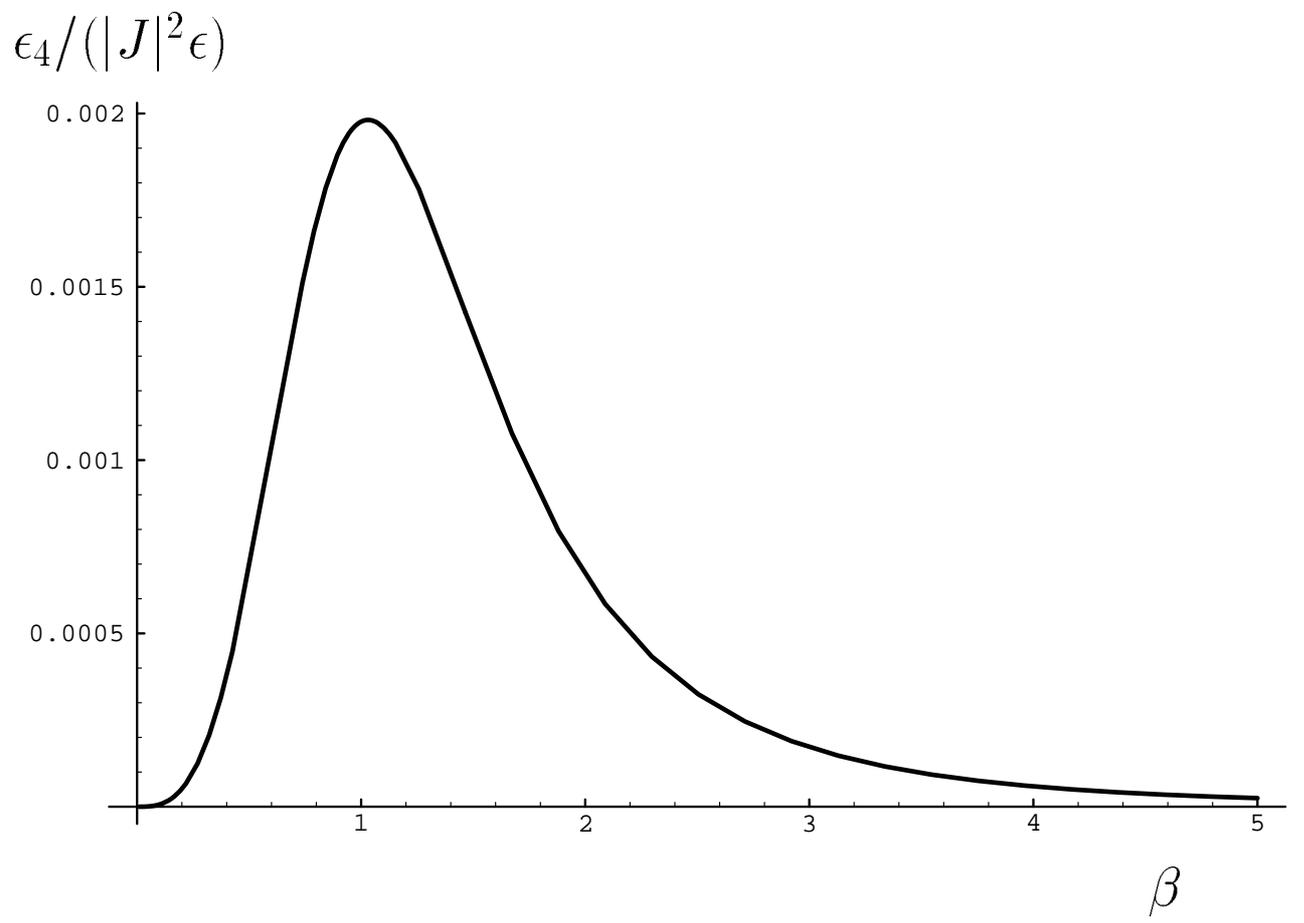
\end{document}